\documentclass[conference]{IEEEtran}

\usepackage{url}
\usepackage{amsmath}

\usepackage[pdftex]{graphicx}
\graphicspath{{_img/}}
\DeclareGraphicsExtensions{.jpeg,.png}

\begin{document}
%
\title{The Generalized Metastable Switch Memristor Model}

\author{
  \IEEEauthorblockN{Timothy W. Molter}
  \IEEEauthorblockA{Knowm Inc., Santa Fe, NM, USA\\
  Email: tim@knowm.org}
  \and
  \IEEEauthorblockN{M. Alexander Nugent}
  \IEEEauthorblockA{Knowm Inc., Santa Fe, NM, USA\\
  Email: alex@knowm.org}
}

\maketitle

\begin{abstract}
Memristor device modeling is currently a heavily researched topic and is becoming ever more important as memristor devices make their way into CMOS circuit designs, necessitating accurate and efficient memristor circuit simulations. In this paper, the Generalized Metastable Switch (MSS) memristor model is presented. The Generalized MSS model consists of a voltage-dependent stochastic component and a voltage-dependent exponential diode current component and is designed to be easy to implement, computationally efficient, and amenable to modeling a wide range of different memristor devices.
\end{abstract}

\IEEEpeerreviewmaketitle

\section{Introduction}

Many memristive materials have recently been reported, and the trend continues. Memristor models are also being developed and incrementally improved upon \cite{biolek2009spice,kvatinsky2015vteam}, however most models to date have shortcomings, especially when used to model the type of devices we are interested in using to build neuromorphic processors such as Thermodynamic-RAM \cite{nugent2014ahah,nugent2014thermodynamic,nugent2014cortical}. Additionally, most memristors seemingly display some measure of stochastic behavior, while most models assume a deterministic device. Naous et al. recently proposed another model, which can add stochasticity to any existing model \cite{naous2015stochasticity}. We felt that a stochastic model built from the ground up was a more natural fit to actual devices and designed it to satisfy several requirements: (1) It should accurately model the device behavior including stochastics, (2) it should be computationally efficient, (3) and it should model as many different devices as possible.∫

\section{Generalized MSS Model}

In our proposed semi-empirical model, the ``generalized metastable switch (MSS) memristor model'', the total current through the device comes from both a memory-dependent current component, $ I_{\rm m} $, and a Schottky diode current, $ I_{\rm s} $ in parallel: 

\begin{equation}I=\phi I_{\rm m}(V,t)+(1-\phi)I_{\rm s}(V)\end{equation} 

, where $ \phi\in{[0,1]} $. A value of $ \phi=1 $ represents a device that contains no Schottky diode effects. The Schottky component, $ I_{\rm s}(V) $, follows from the fact that many memristive devices contain a Schottky barrier formed at a metal--semiconductor junction. The Schottky component is modeled by forward bias and reverse biased components as follows: 

\begin{equation}I_{\rm s}=\alpha_{\rm f}e^{\beta_{\rm f}V}-\alpha_{\rm r}e^{-\beta_{\rm r}V}\end{equation} 

, where $ \alpha_{\rm f} $, $ \beta_{\rm f} $, $ \alpha_{\rm r} $, and $ \beta_{\rm r} $ are positive valued parameters setting the exponential behavior of the forward and reverse exponential current flow across the Schottky barrier. 

The memory component of our model, $ I_{\rm m} $, arises from the notion that memristors can be represented as a collection of conducting channels that switch between 2 states of differing resistance. Modification of device resistance is attained through the application of an external voltage gradient that causes the channels to transition between high and low conducting states. As the number of channels increases, the memristor will become more incremental as it acquires the ability to access more states. We treat each channel as a metastable switch (MSS) and the conductance of a collection of metastable switches captures the memory effect of the memristor. 

The probability that a single MSS will transition from the B state to the A state is given by $P_{\rm A}$, while the probability that the MSS will transition from the $A$ state to the $B$ state is given by $P_{\rm B}$. The transition probabilities are modeled as: 

\begin{equation}P_{\rm A} = \alpha \frac{1}{{1 + {e^{ \beta \left( {V - {V_{\rm A}}} \right)}}}} = \alpha \Gamma \left( { V,{V_{\rm A}}} \right)\end{equation} 

and 

\begin{equation}P_{\rm B} = \alpha \left( {1 - \Gamma \left( {V, -{V_{\rm B}}} \right)} \right)\end{equation} 

, where $ \beta = \frac{q}{{kT}} = {({V_{\rm T}})^{ -1}} $. Here, $ V_{\rm T} $ is the thermal voltage and is equal to approximately $ 26~{{mV}^{-1}} $ at $ T=300~K $, $ \alpha = \frac{{\Delta t}}{{{t_{\rm c}}}} $ is the ratio of the time step period $ \Delta t $ to the characteristic time scale of the device, $ t_{\rm c} $, and $ V $ is the voltage across the device. The probability $ {P_{\rm A}} $ is defined as the positive-going direction, so that a positive applied voltage increases the chances of occupying the A state. Each switch has an intrinsic electrical conductance given by $ G_{\rm A} $ and $ G_{\rm B} $. The convention is that $ G_{\rm B} > G_{\rm A} $. Note that the logistic function $ \frac{1}{{1 + {e^{-x}}}} $ is similar to the hyperbolic-sign function used in other memristive device models. Our use of the logistic function follows simply from the requirement that probabilities must be bounded between 0 and 1. 

Up until this point we have only considered a single MSS being in the $ A $ or $ B $ state and its probability of it changing states given external stimuli. We now model a memristor as a collection of $ N $ MSSs evolving in discrete time steps, $ \Delta t $. The total memristor conductance is given by the sum over each MSS: 

\begin{equation}
{G_{\rm m}} = {N_{\rm A}}{G_{\rm A}} + {N_{\rm B}}{G_{\rm B}}
\end{equation} 

, where $ N_{\rm A} $ is the number of MSSs in the A state, $ N_{\rm B} $ is the number of MSSs in the $ B $ state and $ G_{\rm A} , G_{\rm B} $ are the intrinsic conductances of the MSSs respectively. At each time step some subpopulation of the MSSs in the $ A $ state will transition to the $ B $ state, while some subpopulation in the B state will transition to the A state. Since the model keeps track of the number of switches in each state, it can easily determine the entire device's conductance by a sum of products.

\begin{figure}[!t]
\centering
\includegraphics[width=1.0\linewidth]{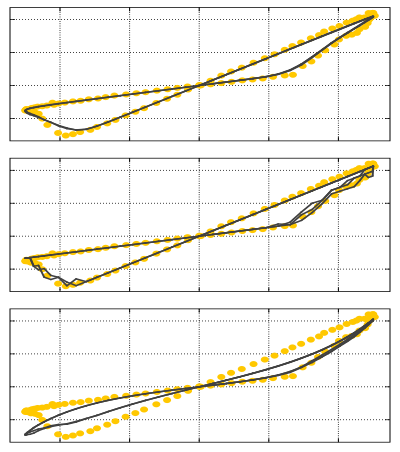}
\caption{The generalized metastable switch model fit to a Tungsten Ag-chalcogenide memristor device. Top) The model parameters are: $N=1000$, $t_{\rm c} = 0.1~ms$, $ G_{\rm A} = 2.125~mS$, $G_{\rm B} = 0.67~mS$, $V_{\rm A} = 0.27~V$, $V_{\rm B} = 0.37~V$ and $\phi = 1$. Center) With $N = 10$, the model is tuned to display more stochastic response. Bottom) With $\phi = .45$, $ \alpha = .00005$, and $\beta = 6$, an exponential behavior response is added.
}
\label{fig_memristor_hyst}
\end{figure}

The probability that $k$ MSSs will transition out of a population of $n$ MSSs is given by the binomial distribution, and as $n$ becomes large we may approximate the binomial distribution with a normal distribution: 

\begin{equation}
\mathcal{N}\left( x \rvert {\mu ,{\sigma ^2}} \right) = \frac{{e^{ - \frac{ {{\left( {x - \mu } \right)}^2}}{{2{\sigma ^2}}}}}}{\sqrt {2\pi {\sigma ^2}} }
\end{equation} 

, where $ \mu = np $ and $ {\sigma ^2} = np\left( {1 - p} \right) $. Therefore at each time step we have two normal distributions defined by $ N_{\rm A}$, $P_{\rm A}$, $ N_{\rm B}$ and $P_{\rm B}$, which is a function of instantaneous voltage, $ V $, the time step, $ \Delta t $, of the simulation and the temperature, $ T $, of the device. These two distributions can be sampled to get a random number of MSSs switching their states, where $ \Delta {N_{\rm A}} $ and $ \Delta {N_{\rm B}} $ are the number of switches transitioning states.

The change in the memristor conductance is thus given by:

\begin{equation}
\Delta G_m = \Delta {N_{\rm A }} \cdot {G_{\rm A}} - \Delta {N_{\rm B}} \cdot {G_{\rm B}}
\end{equation}

, and the memory-dependent current is thus:

\begin{equation}
I_{\rm m}= V \left(G_m + \Delta G_m \right)
\end{equation} 

, where $ V $ is the voltage across the memristor during the time-step. 

\section{Results and Discussion}

Figure \ref{fig_memristor_hyst} shows the hysteresis curve of a fitted model for a raw Tungsten Ag-chalcogenide device data driven at 500 Hz with a sinusoidal voltage of 0.5 V amplitude. Reducing the number of MSSs in the model reduces the averaging effects and causes the memristor to behave in a more stochastic way. Note that as the number of MSSs becomes small, the normal approximation to the binomial distribution also breaks down. In this case, one could use the binomial distribution directly if so desired. An addition of a Schottkey diode current response seen in many memristor devices is illustrated by changing $\phi$, $ \alpha $, and $\beta$.

The generalized metastable switch memristor model presented does an excellent job at modeling the hysteresis behavior of a W-Ag-chalcogenide device with a quick manual fitting procedure. Not shown here is additional satisfactory modelling under a diverse set up simulations: triangle drive waveforms, pulses, positive and negative voltages, two devices connected in series, and additional memristor types. Future work includes porting the model over to Verilog and SPICE. Source code for the model and simulations is available upon request.

\section*{Acknowledgment}
The authors would like to thank the Air Force Research Labs in Rome, NY for their support under the SBIR/STTR programs AF10-BT31, AF121-049. The authors would like to thank Kristy A. Campbell from Boise State University for graciously providing us with memristor device data.

\bibliographystyle{IEEEtran}
\bibliography{2016_MSS}

\end{document}